\documentclass[a4paper,12pt]{article}         
\usepackage{amsmath,amsfonts,amsthm,amssymb}  
\usepackage[utf8]{inputenc} 
\usepackage[T1]{fontenc}    
									   
\DeclareMathOperator*{\argmin}{arg\,min} 

\usepackage{graphicx,caption}            
\usepackage{color}                       
\usepackage{hyperref}                    
\hypersetup{breaklinks=true}

\usepackage{breakurl}

\usepackage{grffile}

\usepackage{makeidx}                  
\usepackage{acronym}
\usepackage{mathrsfs}
\usepackage{float}
\usepackage{vruler}                    
\usepackage{palatino, url, multicol}
\usepackage{setspace}                  
\usepackage{txfonts}

\usepackage{tikz}
\usepackage{tikz-3dplot}
\usetikzlibrary{shapes.geometric, arrows, calc, 3d}

\usepackage[numbers,sort&compress]{natbib}   
\usepackage{import}

\usepackage{caption}
\usepackage{subcaption}

\usepackage{listings}

\usepackage{comment}
\usepackage[title]{appendix}

\restylefloat{figure}

\title{Physically-Based Mesh Generation for Confined 3D Point Clouds Using Flexible Foil Models}
\author{Netzer Moriya}
\date{}

\begin{document}

\maketitle

\begin{abstract}
We propose a method for constructing high-quality, closed-surface meshes from confined 3D point clouds via a physically-based simulation of flexible foils under spatial constraints. 
The approach integrates dynamic elasticity, pressure-driven deformation, and adaptive snapping to fixed vertices, providing a robust framework for realistic and physically accurate mesh creation. Applications in computer graphics and computational geometry are discussed.
\end{abstract}

\section{Introduction}

The problem of mesh generation and deformation has been extensively studied in computational geometry, 
computer graphics, and engineering disciplines. This section provides a brief overview of traditional 
mesh generation techniques, physically-based simulation methods, and applications of elasticity models 
and pressure-driven deformation, with references to key research works in the field.

\subsection{Mesh Generation Techniques}
Mesh generation serves as a foundational step in modeling surfaces and volumes. Among the most widely used techniques are:
\begin{itemize}
    \item \textbf{Delaunay Triangulation:} Delaunay triangulation ensures that the circumcircle of any 
	triangle does not contain other vertices, resulting in well-shaped triangles and numerical stability. 
	Lee and Schachter's seminal work on Delaunay triangulation algorithms laid the groundwork for its 
	computational applications \cite{lee1980two}.
	The method ensures that the circumcircle of any triangle does not contain other vertices, leading to 
	well-shaped triangles and numerical stability. It can struggle with concave regions 
	as it prioritizes triangle shape over accurate concavity capture.

    \item \textbf{Poisson Surface Reconstruction:} Proposed by Kazhdan et al. \cite{kazhdan2006poisson}, 
	this method generates meshes from point cloud data by solving a screened Poisson equation, enabling 
	smooth and watertight surface reconstructions. It has been extensively used in reverse engineering and 
	3D scanning applications.
    The Poisson Surface Reconstruction excels at handling concavities and produces smooth, however it does 
	suffer from several drawbacks:
	\begin{itemize}
        \item \textbf{Assumes Watertight Surfaces:} PSR assumes watertight surfaces, making it unsuitable for open 
		or perforated objects.
        \item \textbf{Sensitive to Noise:} Noisy point clouds can introduce artifacts in the reconstructed surface.
        \item \textbf{Loss of Detail:} Fine details and high-curvature areas may be smoothed out.
        \item \textbf{No Sharp Features:} Struggles to preserve sharp edges and corners, producing smooth surfaces 
		instead.
        \item \textbf{Normals Dependency:} Relies on accurate and oriented normals; errors lead to incorrect 
		surfaces.
        \item \textbf{Computation Time:} Reconstruction can be slow for large datasets or high precision.
    \end{itemize}
	
    \item \textbf{Convex Hulls and Voronoi Diagrams:} Convex hull algorithms, such as those by Preparata 
	and Shamos \cite{preparata1985computational}, are crucial for creating initial meshes and understanding 
	spatial relationships in 3D domains.
    Inherently however, the method cannot represent concavities by definition.

    \item \textbf{Ball-Pivoting Algorithm (BPA):} 
    This method "rolls" a ball over the point cloud, connecting points to form triangles. It's simple 
	and efficient for dense data but may miss narrow concavities.

    \item \textbf{Alpha Shapes:} 
    Controlled by a parameter $\alpha$, alpha shapes capture the overall form of a point cloud, 
	including concavities. Although widely used \cite{Edelsbrunner1987, MATLAB}, this approach is useful 
	for filtering noise and outliers but can lead to non-watertight, open surface artifacts.

    \item \textbf{Marching Cubes:} 
    While not directly applied to point clouds, this algorithm generates meshes from volumetric data. 
	It's often used in conjunction with techniques that convert point clouds to volumes.

    \item \textbf{Advancing Front Methods:} 
    These methods progressively add triangles to a mesh, starting from an initial boundary. 
	They can handle complex shapes and concavities effectively however:
    \begin{itemize}
        \item \textbf{Sensitive to Initialization:} The quality of the mesh heavily depends on the initial 
		boundary and seed points.
        \item \textbf{Complexity in Implementation:} Implementing advancing front methods can be challenging 
		due to intricate boundary and front management.
        \item \textbf{Poor Handling of Large Data:} These methods may struggle with scalability and efficiency 
		when dealing with very large point clouds.
        \item \textbf{Difficulty with Noisy Data:} Sensitive to noise, which can lead to poorly formed 
		triangles or errors in the front propagation.
        \item \textbf{Topological Challenges:} Handling topological changes, such as holes or complex 
		connections, can be difficult.
        \item \textbf{Non-Uniform Sampling Issues:} Regions with non-uniform point sampling may result 
		in irregular or distorted mesh elements.
    \end{itemize}
		
\end{itemize}

These methods form the basis of modern approaches to high-quality mesh generation, often enhanced with adaptive refinement, smoothing, and hierarchical representations \cite{botsch2010polygon}.

\subsection{Physically-Based Simulation Methods}
Physically-based simulation techniques, rooted in continuum mechanics, model the behavior of deformable objects under external forces, 
grounded in continuum mechanics. Key approaches include:

\begin{itemize}
    \item \textbf{Mass-Spring Systems:} Widely used for real-time simulations, mass-spring systems discretize 
	an object into a network of masses connected by springs, with elasticity modeled using Hooke’s Law. 
	Provot \cite{provot1995deformation} demonstrated the effectiveness of these systems in cloth simulation, 
	though challenges remain in accurately handling large deformations.
    \item \textbf{Finite Element Method (FEM):} FEM provides a mathematically rigorous framework for simulating 
	deformable bodies by discretizing the governing equations over finite elements. 
	Zienkiewicz et al. \cite{zienkiewicz1977finite} offer a comprehensive treatment of FEM in engineering 
	contexts, while Terzopoulos et al. \cite{terzopoulos1987elastically} adapted FEM for 
	physically-based modeling in graphics.
    \item \textbf{Position-Based Dynamics (PBD):} Introduced by Müller et al. \cite{muller2007position}, 
	PBD provides an efficient alternative to traditional force-based methods by solving for positions 
	directly, making it suitable for interactive applications, however, may not always capture material properties 
	accurately.
    \item \textbf{Smoothed Particle Hydrodynamics (SPH):} SPH \cite{gingold1977smoothed} is a meshfree method that represents objects as a 
	set of interacting particles. It's particularly well-suited for simulating fluids and highly deformable 
	materials.

    \item \textbf{Boundary Element Method (BEM):}
    BEM focuses on the boundary of the object \cite{brebbia1984boundary}, making it computationally efficient for problems with complex 
	geometries. However, it can be challenging to implement for non-linear materials.

    \item \textbf{Discrete Element Method (DEM):}
    DEM models objects as a collection of discrete elements interacting through contact forces \cite{cundall1979discrete}. 
	It's commonly used for granular materials and fracture simulations.

\end{itemize}

Modern simulation frameworks often combine these approaches, leveraging their respective strengths to achieve 
realism and computational efficiency \cite{sifakis2012fem, BaraffWitkin1998}.

\subsection{Applications of Elasticity Models and Pressure-Driven Deformation}
Elasticity models play a critical role in simulating deformable objects, particularly under pressure forces. Pressure-driven deformation has been studied in various contexts:
\begin{itemize}
    \item \textbf{Biomedical Applications:} Elastic models are used to simulate soft tissue deformation for surgical planning and medical training. A notable example is the work by Taylor et al. \cite{taylor2008bio}, which integrates biomechanical simulations with imaging data~\cite{TeranEtAl2005}.
    \item \textbf{Aeroelasticity and Fluid-Structure Interaction:} In engineering, elasticity and pressure forces are critical in understanding how flexible structures interact with surrounding fluid flows. The works of Bathe \cite{bathe1996fluid} provide foundational insights into coupling structural mechanics with fluid dynamics.
    \item \textbf{Computer Graphics:} Elasticity-driven deformations have applications in creating realistic animations and special effects, as demonstrated by Debunne et al. \cite{debunne2001dynamic}, who introduced real-time methods for simulating large deformations.
\end{itemize}

Recent advancements in this domain focus on adaptive elasticity models that incorporate strain-dependent stiffness and proximity-based corrections, ensuring stability in highly dynamic scenarios \cite{wang2015adaptive}.

While existing methods in mesh generation, physically-based simulation, and elasticity-driven deformation have 
made significant strides in diverse applications, they often focus on either static configurations or generic 
deformation scenarios. 
In contrast, our approach uniquely models the dynamics of a flexible foil subjected to contractive pressure 
while constrained by fixed vertices and governed by mesh elasticity. 
This formulation not only captures the interplay between external forces and internal elastic behavior but 
also introduces adaptive constraints that ensure geometric fidelity and stability throughout the deformation 
process. By integrating principles from computational geometry, continuum mechanics, and numerical simulation, 
our method addresses key challenges in achieving physically accurate and computationally efficient deformation 
modeling. This work extends the applicability of mesh-based simulations to scenarios requiring precise control 
under complex constraints, bridging the gap between foundational research and advanced applications in areas 
such as engineering, biomedical modeling, and computer graphics.

Building upon our earlier work \cite{netzer2024form}, which introduced the \emph{Point Cloud Contraction Theorem}, 
this study extends its application to dynamic simulations of deformable meshes. 
The theorem establishes a formal mechanism for transforming a convex hull into 
a concave surface, ensuring the resulting geometry encloses all points of the underlying dataset 
while maintaining a closed, non-intersecting structure. Its iterative facet replacement and 
expansion process provides a robust mathematical framework for modeling contraction phenomena.
In this work, we leverage these principles to develop a pressure-driven deformation model, constrained by 
fixed vertices and governed by mesh elasticity. Specifically, the adaptive deformation mechanism directly 
incorporates insights from the theorem, ensuring that the mesh evolves in a physically consistent manner, 
balancing contractive forces and elastic constraints. This advancement bridges the theoretical foundations 
of surface contraction with practical applications in computational geometry, elasticity modeling, and dynamic 
simulation frameworks for constrained deformations. By integrating these principles, we 
demonstrate the versatility and utility of the Point Cloud Contraction Theorem in addressing real-world 
challenges in pressure-driven systems.

\section{Mathematical Formulation}
\subsection{Problem Definition}
The objective of this study is to model the deformation of a flexible foil mesh under the influence of external and internal forces. Mathematically, the problem is defined over a discretized domain represented as a triangular mesh $\mathcal{M} = (\mathcal{V}, \mathcal{F})$, where:
\begin{itemize}
    \item $\mathcal{V} = \{ \mathbf{v}_i \in \mathbb{R}^3 : i = 1, \dots, n \}$ is the set of $n$ vertices with positions $\mathbf{v}_i$ in 
	three-dimensional space. The vertices $\mathbf{v}_i$, are assumed to constraint the mesh during its evolvement toward final shape.
    \item $\mathcal{F} \subseteq \mathcal{V}^3$ is the set of triangular faces formed by the vertices in $\mathcal{V}$.
\end{itemize}

The deformation process involves determining a new configuration $\mathcal{V}' = \{ \mathbf{v}_i' \}$ of the vertices such that equilibrium is achieved under the combined effects of elastic and pressure forces, subject to the following constraints:
\begin{enumerate}
    \item \textbf{Elastic Force Equilibrium:} The forces $\mathbf{f}_i^{\text{elastic}}$ are derived from Hooke's law, ensuring that the edge lengths approach their rest lengths $L_{ij}$:
    \begin{equation}
    \mathbf{f}_i^{\text{elastic}} = \sum_{j \in \mathcal{N}(i)} k_{ij} \left( \frac{\|\mathbf{v}_j - \mathbf{v}_i\| - L_{ij}}{L_{ij}} \right) \frac{\mathbf{v}_j - \mathbf{v}_i}{\|\mathbf{v}_j - \mathbf{v}_i\|},
    \end{equation}
    where $k_{ij}$ is the stiffness coefficient, $\mathcal{N}(i)$ is the set of neighboring vertices of $\mathbf{v}_i$, and $\|\cdot\|$ denotes the Euclidean norm.

    \item \textbf{Pressure Force Equilibrium:} The forces $\mathbf{f}_i^{\text{pressure}}$ are computed based on the local or global pressure acting on the mesh surface. For a triangular face $f = (\mathbf{v}_i, \mathbf{v}_j, \mathbf{v}_k)$, the force contribution is:
    \begin{equation}
    \mathbf{f}_i^{\text{pressure}} = \frac{p \cdot \mathbf{n}_f \cdot A_f}{3},
    \end{equation}
    where $p$ is the pressure magnitude, $\mathbf{n}_f$ is the unit normal vector of the face, and $A_f$ is 
	the area of the triangle. In this formulation, we assume uniform pressure throughout the mesh. 
	However, this assumption can be adjusted to accommodate specific simulation conditions or application 
	requirements by incorporating spatially varying pressure distributions.

    \item \textbf{Geometric Constraints and Snapping:} Certain vertices $\mathcal{V}_{\text{fixed}} \subseteq \mathcal{V}$ are designated as fixed, enforcing the positional constraints $\mathbf{v}_i' = \mathbf{v}_i, \, \forall \mathbf{v}_i \in \mathcal{V}_{\text{fixed}}$. Additionally, snapping ensures that each vertex $\mathbf{x}$ is projected 
	to the nearest fixed vertex, minimizing the Euclidean distance:
    \begin{equation}
    \mathbf{d}_{\text{snap}} = \min_{\mathbf{v}_i \in \mathcal{V}_{\text{fixed}}} \| \mathbf{x} - \mathbf{v}_i \|.
    \label{eq:snapp}
    \end{equation}

    \item \textbf{Smoothing:} To maintain mesh quality and stability, vertices are periodically adjusted to achieve a locally consistent configuration relative to the deformed mesh. Smoothing tolerances ensure that vertices remain within $\epsilon$ of their intended positions, preserving geometric regularity without assuming a predefined global shape.

\end{enumerate}

The governing equation for the dynamic system is expressed as:

\begin{equation}
m_i \frac{d^2 v_i}{dt^2} = f_{\text{elastic}, i} + f_{\text{pressure}, i} - c \frac{dv_i}{dt},
\label{eq:motion}
\end{equation}

where:
\begin{itemize}
    \item \( m_i \) is the mass associated with vertex \( v_i \),
    \item \( f_{\text{elastic}, i} \) is the elastic force exerted by neighboring vertices,
    \item \( f_{\text{pressure}, i} \) is the pressure-induced force acting on the vertex,
    \item \( c \) is the damping coefficient, which introduces a velocity-dependent resistive force \( -c \frac{dv_i}{dt} \).
\end{itemize}

Equation~\eqref{eq:motion} describes the motion of each vertex in the flexible foil mesh under 
the influence of elastic forces, pressure forces, and damping. This equation is derived from 
Newton’s Second Law of Motion:

This formulation represents a damped mass-spring system, where:
\begin{itemize}
    \item The \textbf{elastic term} models restoring forces that maintain structural integrity.
    \item The \textbf{pressure term} simulates external forces deforming the mesh.
    \item The \textbf{damping term} prevents excessive oscillations and ensures numerical stability.
\end{itemize}

\subsubsection{Role of Damping \( c \)}
The damping coefficient \( c \) is introduced to control undesirable oscillations in the mesh. It serves two key purposes:
\begin{enumerate}
    \item \textbf{Numerical Stability:} In explicit integration methods (such as Euler), undamped oscillations can lead to instability. Proper damping reduces high-frequency vibrations.
    \item \textbf{Physical Realism:} In real-world materials, energy dissipation occurs through internal friction or air resistance. The term \( -c \frac{dv_i}{dt} \) mimics this effect.
\end{enumerate}

The choice of \( c \) affects the system's behavior:
\begin{itemize}
    \item \textbf{Underdamping} (\( c \) too small): The system oscillates excessively before settling.
    \item \textbf{Critical damping} (\( c_{\text{crit}} = 2\sqrt{k m} \)): The system reaches equilibrium in minimal time without oscillations.
    \item \textbf{Overdamping} (\( c > c_{\text{crit}} \)): Motion is slowed down but remains stable.
\end{itemize}

In our implementation, \( c \) is chosen to be close to or slightly above the critical damping value 
to ensure fast and stable convergence without excessive oscillations.

The numerical simulation employs an explicit Euler integration scheme with a time step \( \Delta t \). 
At each iteration, the vertex positions and velocities are updated based on the computed forces. 
The process continues until a convergence criterion—such as a maximum displacement threshold or an 
energy stabilization condition—is satisfied.

\subsection{Dynamic Simulation}
The dynamic simulation of the flexible foil mesh evolves the positions and velocities of vertices over discrete time steps. The process employs an explicit Euler integration scheme, governed by the following equations:
\begin{align}
\mathbf{v}_{t+1} &= \mathbf{v}_t + \Delta t \cdot \mathbf{a}_t, \label{eq:euler_velocity} \\
\mathbf{x}_{t+1} &= \mathbf{x}_t + \Delta t \cdot \mathbf{v}_{t+1}, \label{eq:euler_position}
\end{align}
where:
\begin{itemize}
    \item $\mathbf{v}_t$ and $\mathbf{v}_{t+1}$ are the velocities of the vertices at times $t$ and $t+1$, respectively.
    \item $\mathbf{x}_t$ and $\mathbf{x}_{t+1}$ are the vertex positions at times $t$ and $t+1$.
    \item $\Delta t$ is the integration time step, selected to ensure numerical stability.
    \item $\mathbf{a}_t = \frac{\mathbf{f}_t}{m}$ is the acceleration vector at time $t$, derived from the net force $\mathbf{f}_t$ acting on a vertex and its mass $m$.
\end{itemize}

The total force $\mathbf{f}_t$ is the sum of contributions from elastic forces, pressure forces, and damping effects:
\begin{equation}
\mathbf{f}_t = \mathbf{f}_t^{\text{elastic}} + \mathbf{f}_t^{\text{pressure}} - c \cdot \mathbf{v}_t,
\end{equation}
where $c$ is the damping coefficient used to stabilize the system by mitigating oscillatory behavior.

\subsubsection{Convergence Criteria}
To determine when the simulation has reached equilibrium, a convergence criterion is imposed on the maximum displacement of the vertices between successive time steps:
\begin{equation}
\max(\| \mathbf{x}_{t+1} - \mathbf{x}_t \|) < \epsilon, \label{eq:convergence}
\end{equation}
where $\epsilon$ is the displacement tolerance. When this condition is satisfied, the simulation halts, indicating that the deformation has stabilized.

\subsubsection{Stability Considerations}

To ensure numerical stability and accuracy in the simulation, two key factors must be considered: 
\textbf{damping effects} and \textbf{time-step constraints}. The explicit Euler method, while computationally 
efficient, is known to be conditionally stable, requiring careful parameter selection.

\paragraph{Damping Effects and System Stability}

Damping plays a crucial role in stabilizing the simulation by reducing high-frequency oscillations and 
preventing numerical divergence. The damping coefficient \( c \) in Equation~\eqref{eq:motion} introduces a 
velocity-proportional resistive force, ensuring that the system does not exhibit excessive oscillatory behavior. 
In particular, damping helps to:
\begin{itemize}
    \item Suppress unwanted high-frequency oscillations in the mesh.
    \item Improve convergence speed by reducing the energy of unstable modes.
    \item Prevent instabilities that arise in explicit integration schemes when forces fluctuate rapidly.
\end{itemize}

For optimal stability, the damping coefficient is typically chosen to be close to or slightly above the critical 
damping threshold \( c_{\text{crit}} = 2\sqrt{k m} \), where \( k \) represents the effective stiffness and \( m \) 
the mass of each vertex. Overdamping (\( c > c_{\text{crit}} \)) slows down the motion, while underdamping 
(\( c < c_{\text{crit}} \)) may allow oscillations to persist.

\paragraph{Time-Step Restrictions: The CFL Condition}

In addition to damping, the simulation time step \( \Delta t \) must satisfy the Courant-Friedrichs-Lewy (CFL) 
condition \cite{Courant_1928, Strikwerda_2004} to prevent numerical instability. The CFL condition imposes an upper 
bound on \( \Delta t \), given by:

\begin{equation}
\Delta t < \frac{2}{\omega_{\text{max}}},
\label{eq:CFL}
\end{equation}

where \( \omega_{\text{max}} \) is the highest eigenfrequency of the system. This frequency is determined by the 
stiffness properties of the mesh and the mass distribution of the vertices.

If \( \Delta t \) is too large, numerical errors accumulate, leading to unstable behavior and divergence. By 
ensuring compliance with the CFL condition, the explicit Euler integration method remains stable and accurately 
tracks the system’s dynamics.

\subsubsection{Computational Workflow}
At each time step, the simulation proceeds as follows:
\begin{enumerate}
    \item Compute the net forces $\mathbf{f}_t$ for all vertices, including contributions from elastic and pressure forces.
    \item Update accelerations $\mathbf{a}_t$, velocities $\mathbf{v}_{t+1}$, and positions $\mathbf{x}_{t+1}$ using Equations \eqref{eq:euler_velocity} and \eqref{eq:euler_position}.
    \item Check the convergence criterion in Equation \eqref{eq:convergence}. If satisfied, terminate the simulation.
    \item Apply optional mesh smoothing and snapping constraints to maintain geometric quality.
    \item Proceed to the next time step.
\end{enumerate}

The explicit Euler integration method, while computationally straightforward, requires careful parameter tuning to ensure stability and to accurately capture the physical behavior of the system. The approach balances computational efficiency with the fidelity of the simulation results.

\section{Implementation Details}
The implementation of the simulation is structured to ensure high-quality mesh generation, geometric accuracy, and computational efficiency. The following subsections detail the key components of the methodology.

\subsection{Mesh Initialization}
The initial mesh $\mathcal{M} = (\mathcal{V}, \mathcal{F})$ is generated using a Fibonacci 
lattice \cite{aistleitner2012, extremelearning}, a method known for producing uniformly distributed points on a sphere. 
Given a target resolution parameter $N$, the vertices $\mathcal{V}$ are computed as:
\begin{align}
\phi_i &= \arccos\left(1 - \frac{2(i + 0.5)}{N}\right), \\
\theta_i &= \pi \cdot (1 + \sqrt{5}) \cdot i, \\
\mathbf{v}_i &= \begin{bmatrix}
\sin(\phi_i) \cos(\theta_i) \\
\sin(\phi_i) \sin(\theta_i) \\
\cos(\phi_i)
\end{bmatrix} \cdot R,
\end{align}
where $R$ is the radius of the sphere, and $i = 1, \dots, N$ indexes the points. This formulation ensures an approximately 
equal angular spacing between the points.

\subsection{Initial Mesh Generation Using Convex Hull}
Once the vertices $\mathcal{V}$ are defined, the triangular faces $\mathcal{F}$ are constructed using the convex hull 
of the vertices. Mathematically, the convex hull is defined as:
\begin{equation}
\text{Conv}(\mathcal{V}) = \bigcap_{H \supseteq \mathcal{V}} H,
\end{equation}
where $H$ is a convex set containing $\mathcal{V}$. The faces $\mathcal{F}$ are extracted as the triangular simplices from the convex hull, ensuring a watertight and topologically consistent mesh.

\subsection{Incorporation of Fixed Vertices and Snapping}
A subset of fixed vertices $\mathcal{V}_{\text{fixed}} \subseteq \mathcal{V}$ is incorporated into the mesh to enforce positional constraints. The snapping mechanism aligns vertices near fixed positions to the nearest fixed vertex. This process is mathematically expressed as:
\begin{equation}
\mathbf{x}_i' = \argmin_{\mathbf{v}_j \in \mathcal{V}_{\text{fixed}}} \|\mathbf{x}_i - \mathbf{v}_j\|, \quad \forall \mathbf{x}_i \in \mathcal{V}.
\end{equation}
Snapping is performed iteratively, and neighboring vertices are adjusted using a relaxation method to ensure smooth transitions between constrained and unconstrained regions.

\subsection{Iterative Smoothing and Refinement}
To maintain geometric regularity and improve mesh quality, iterative smoothing is applied to the mesh (see also in~\cite{AmentaBern1999}). 
A Laplacian smoothing operator is employed, defined as:
\begin{equation}
\mathbf{x}_i' = \mathbf{x}_i + \lambda \sum_{j \in \mathcal{N}(i)} (\mathbf{x}_j - \mathbf{x}_i),
\end{equation}
where $\mathcal{N}(i)$ is the set of neighboring vertices of $\mathbf{x}_i$, and $\lambda \in (0, 1)$ is the smoothing parameter. To prevent distortion near fixed vertices, the smoothing operation is constrained to exclude fixed vertices:
\begin{equation}
\mathbf{x}_i' = \mathbf{x}_i, \quad \text{if } \mathbf{x}_i \in \mathcal{V}_{\text{fixed}}.
\end{equation}

Refinement is achieved by mesh subdivision, wherein each triangular face is subdivided into smaller triangles to enhance resolution. The subdivision process can be expressed recursively:
\begin{equation}
\mathcal{F}' = \bigcup_{f \in \mathcal{F}} \text{subdivide}(f),
\end{equation}
where $\text{subdivide}(f)$ splits a triangle into smaller triangles by introducing midpoints along its edges.

\subsection{Iterative Refinement and Validation}
Each iteration of the refinement process includes:
\begin{itemize}
    \item \textbf{Validation of Mesh Quality:} The mesh is checked for self-intersections and degeneracies. A quality metric, such as the minimum angle $\theta_{\text{min}}$ of each triangle, is used:
    \begin{equation}
    \theta_{\text{min}}(f) = \min \{\theta_1, \theta_2, \theta_3\}, \quad f \in \mathcal{F}.
    \end{equation}
    \item \textbf{Geometric Adjustment:} After refinement, vertices are adjusted to maintain geometric accuracy by aligning with local mesh consistency or predefined constraints:
    \begin{equation}
    \mathbf{x}_i' = \mathbf{c} + R \cdot \frac{\mathbf{x}_i - \mathbf{c}}{\|\mathbf{x}_i - \mathbf{c}\|}.
    \end{equation}
	where $c$ and $R$ refer to the center and radius of the initial sphere respectively.
\end{itemize}

This iterative refinement ensures that the mesh remains suitable for dynamic simulation while adhering to the geometric constraints imposed by the problem definition.

\subsection{Implementation Workflow}
The overall implementation workflow is as follows:
\begin{enumerate}
    \item Generate an initial spherical mesh using the Fibonacci lattice.
    \item Compute the convex hull to form the triangular faces of the mesh.
    \item Incorporate fixed vertices and perform snapping to enforce constraints.
    \item Apply iterative smoothing and refinement to improve geometric regularity.
    \item Validate the refined mesh and adjust vertex positions to maintain geometric consistency relative to the deformed configuration.
\end{enumerate}

The proposed implementation balances computational efficiency with geometric precision, ensuring a robust foundation for subsequent dynamic simulations.

\section{Results and Evaluation}

In the following, we present the implementation of the Flexible Foil mesh on a simple 3D structure. 
The test scenario consists of a symmetrical box where four fixed vertices are positioned at the centers of 
the box's inner faces and slightly inside the outer surface of the box (85\% inside), serving as constraints for the evolving mesh.

\begin{figure}[H]
    \centering
    \begin{subfigure}[t]{0.48\textwidth}
        \centering
        \includegraphics[width=\textwidth]{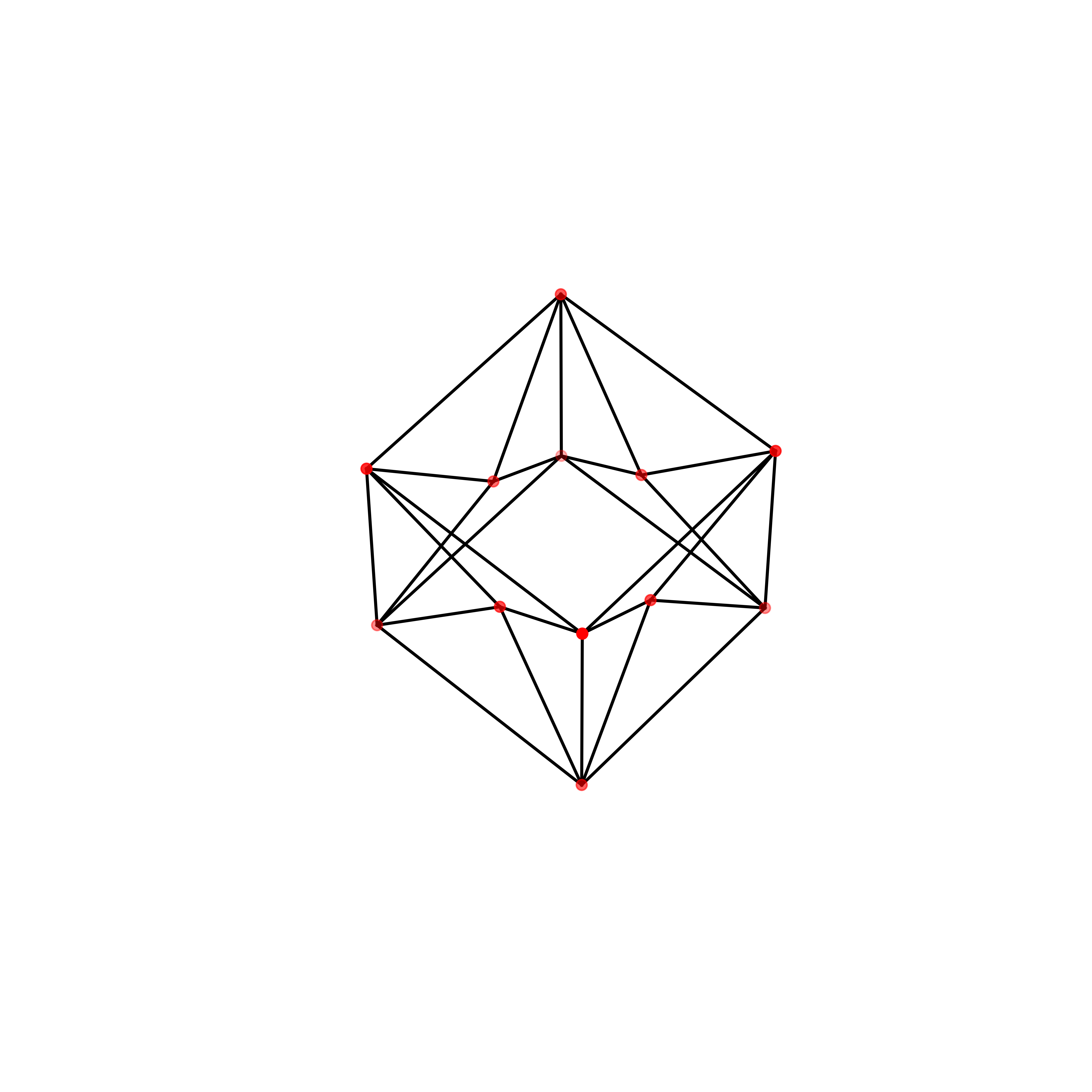}
        \caption{3D structure with red points referring to fixed vertices and black lines to guide the eye.}
    \end{subfigure}
    \hfill
    \begin{subfigure}[t]{0.48\textwidth}
        \centering
        \includegraphics[width=\textwidth]{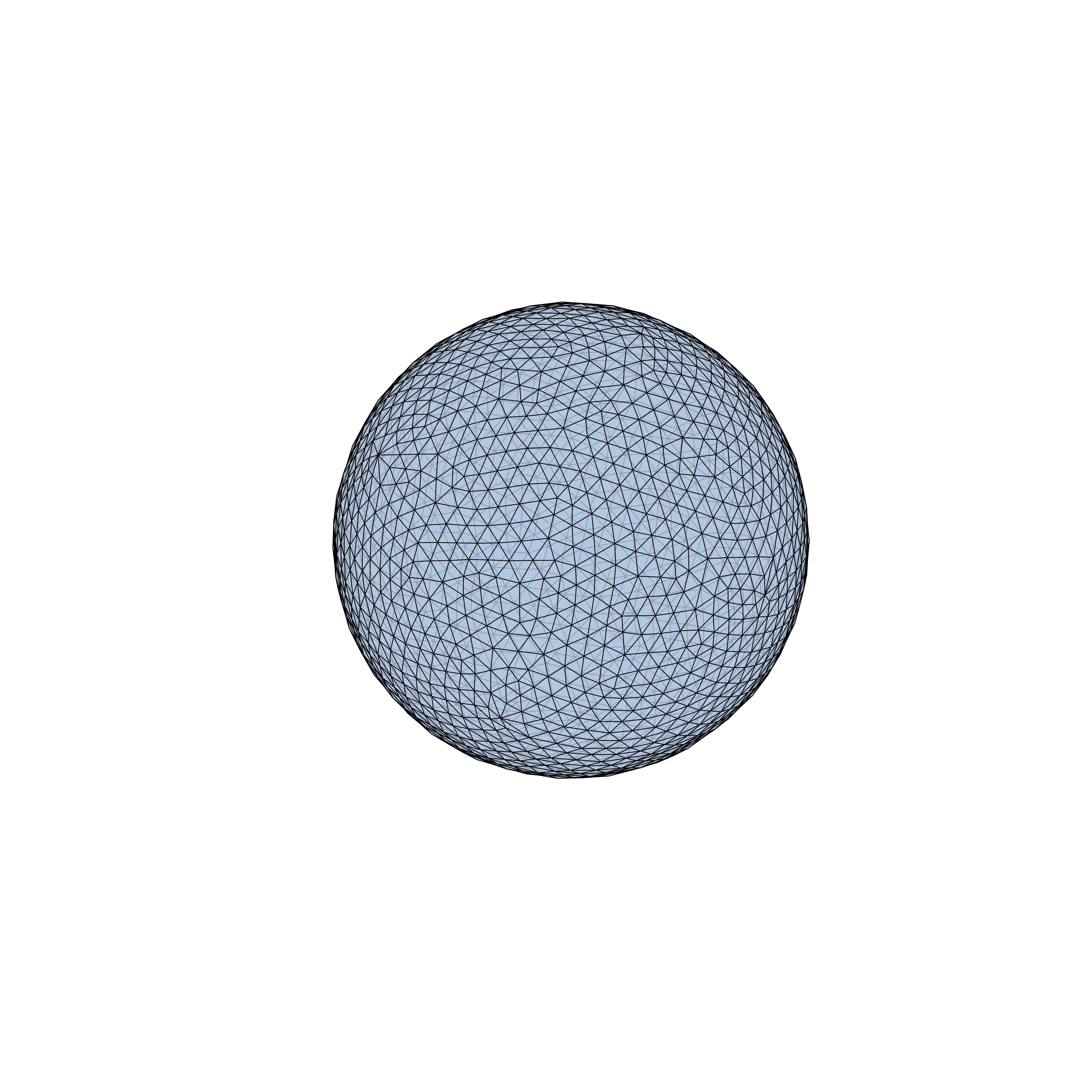}
        \caption{Enclosing sphere and initial mesh definition.}
    \end{subfigure}
    \caption{(Left) Fixed vertices and guiding lines. (Right) Sphere and initial mesh definition.}
    \label{fig:vertices_lines_sphere}
\end{figure}

The initial configuration is shown in Figure~\ref{fig:vertices_lines_sphere}, where the enclosing 
sphere defines the initial mesh structure.

\subsection{Mesh Evolution and Convergence Behavior}

\begin{figure}[H]
    \centering
    \begin{subfigure}[t]{0.48\textwidth}
        \centering
        \includegraphics[width=\textwidth]{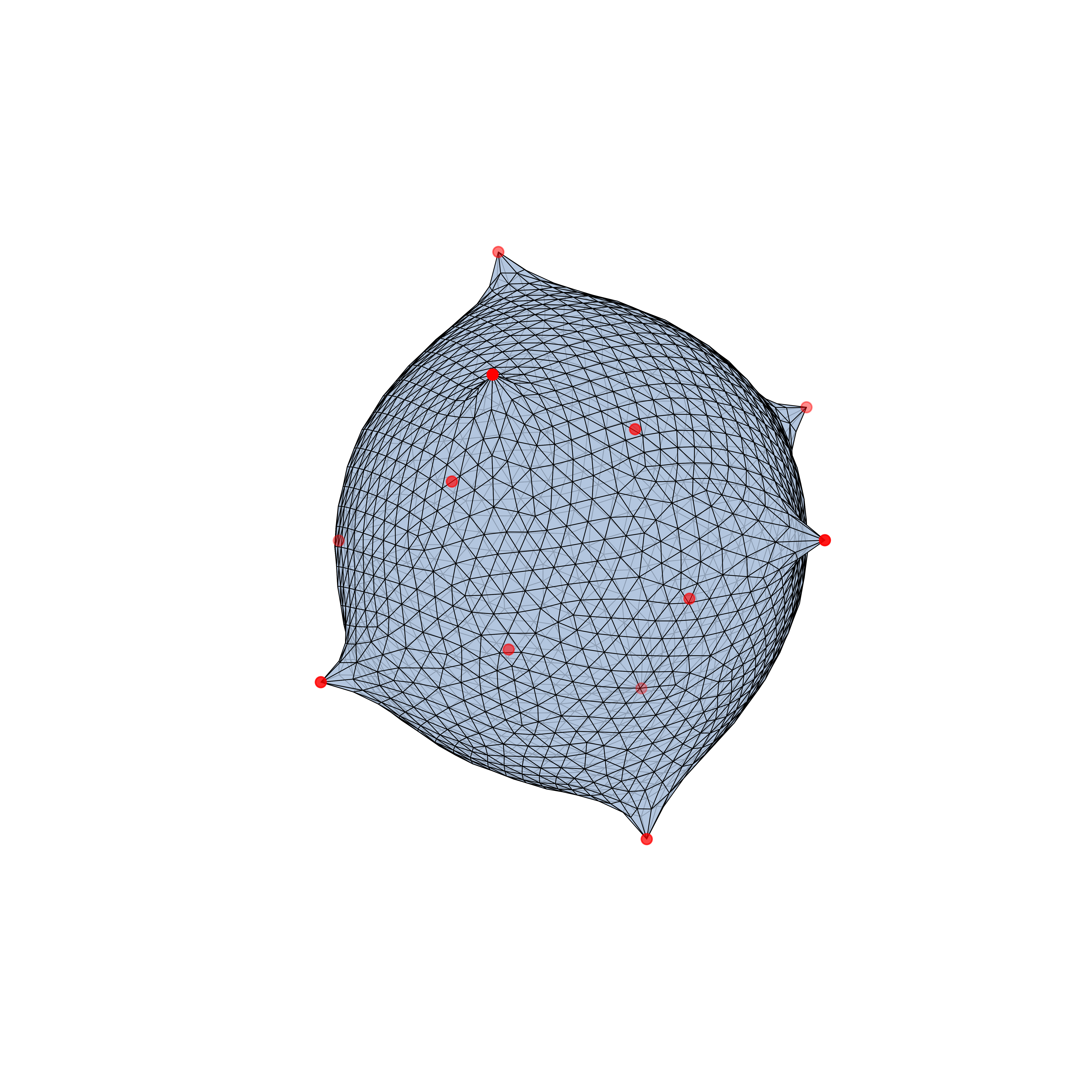}
        \caption{Mesh after 10 iterations.}
    \end{subfigure}
    \hfill
    \begin{subfigure}[t]{0.48\textwidth}
        \centering
        \includegraphics[width=\textwidth]{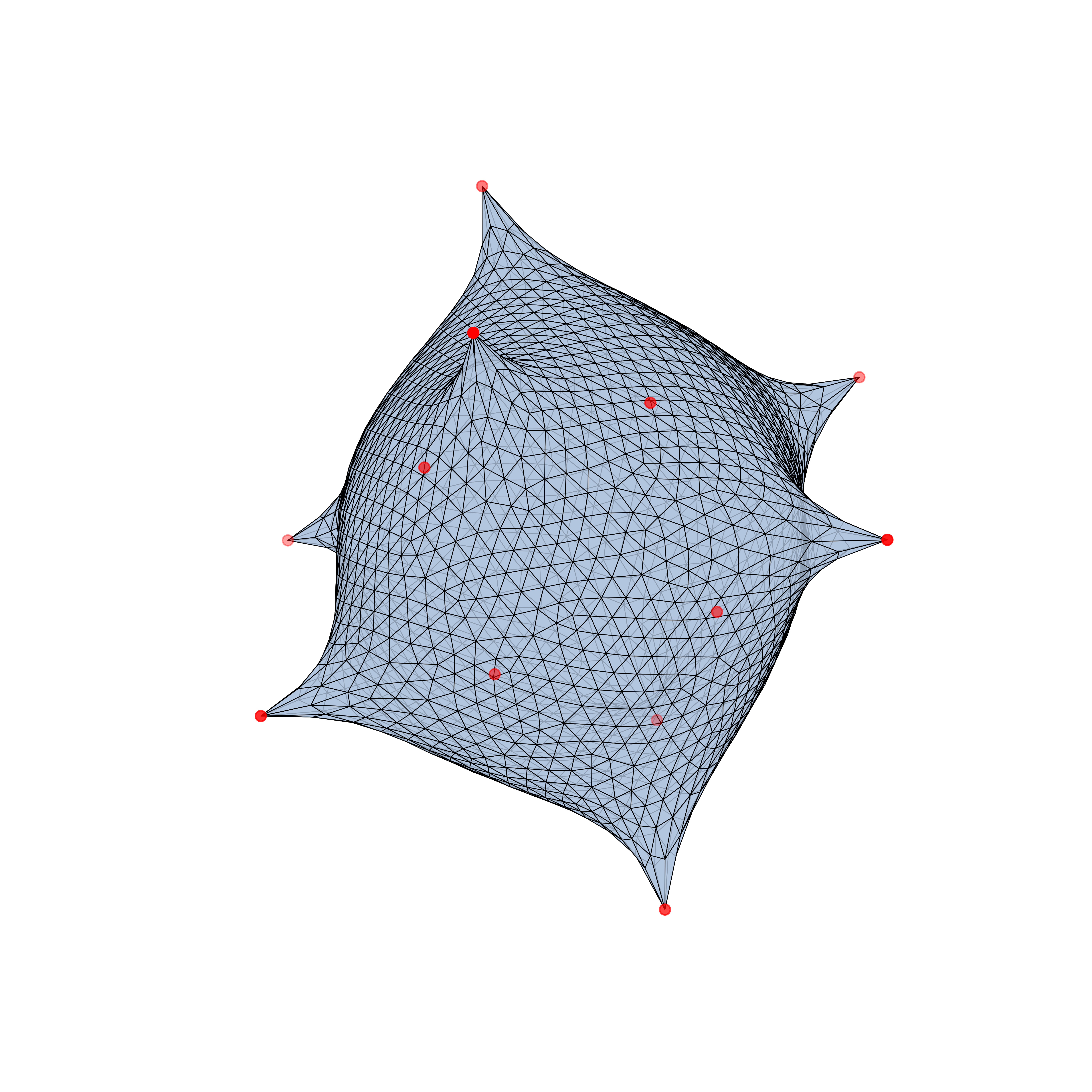}
        \caption{Mesh after 20 iterations.}
    \end{subfigure}
    \caption{Mesh evolution after 10 and 20 iterations, showing initial contraction and alignment.}
    \label{fig:10_20_iterations}
\end{figure}

\begin{figure}[H]
    \centering
    \begin{subfigure}[t]{0.48\textwidth}
        \centering
        \includegraphics[width=\textwidth]{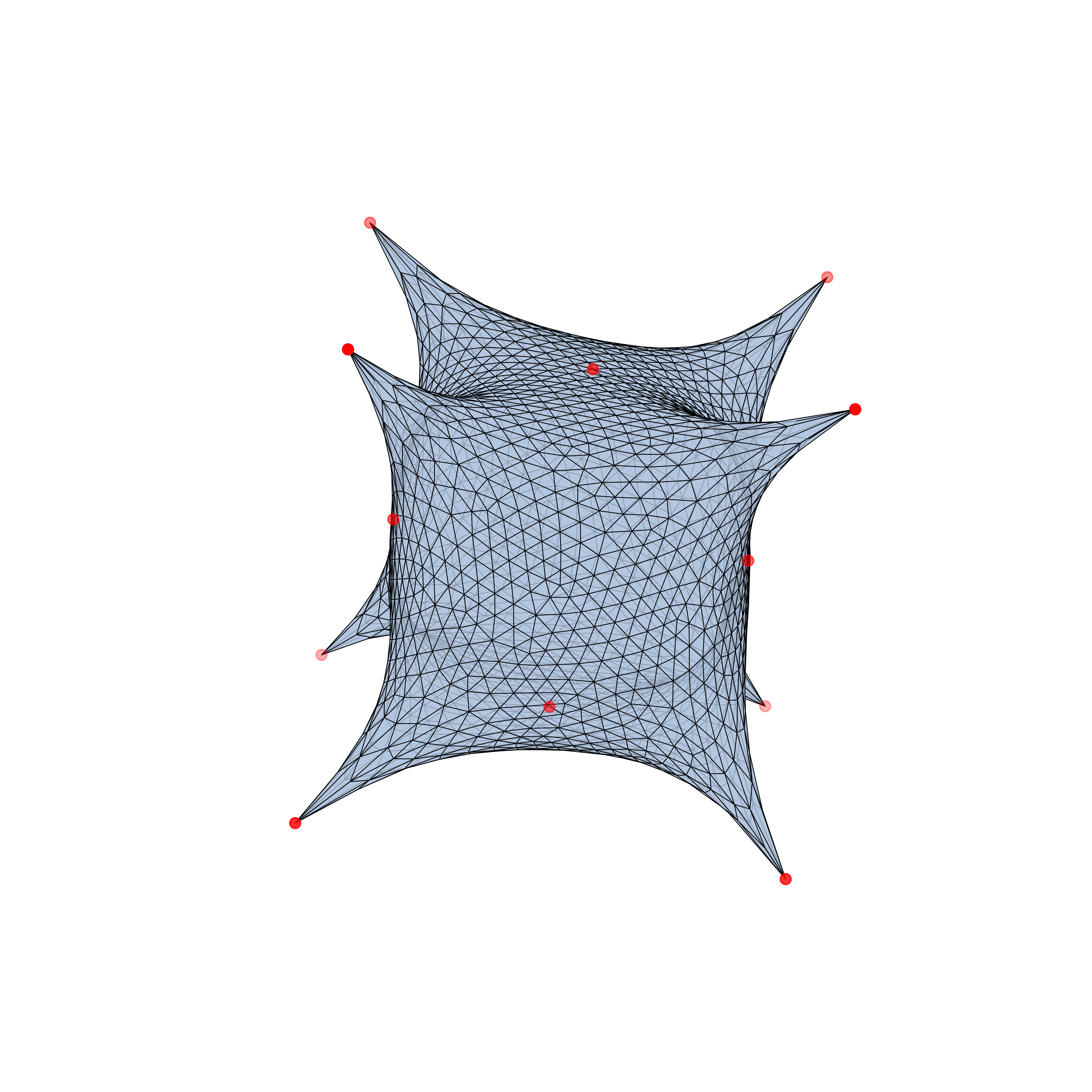}
        \caption{Mesh after 40 iterations.}
    \end{subfigure}
    \hfill
    \begin{subfigure}[t]{0.48\textwidth}
        \centering
        \includegraphics[width=\textwidth]{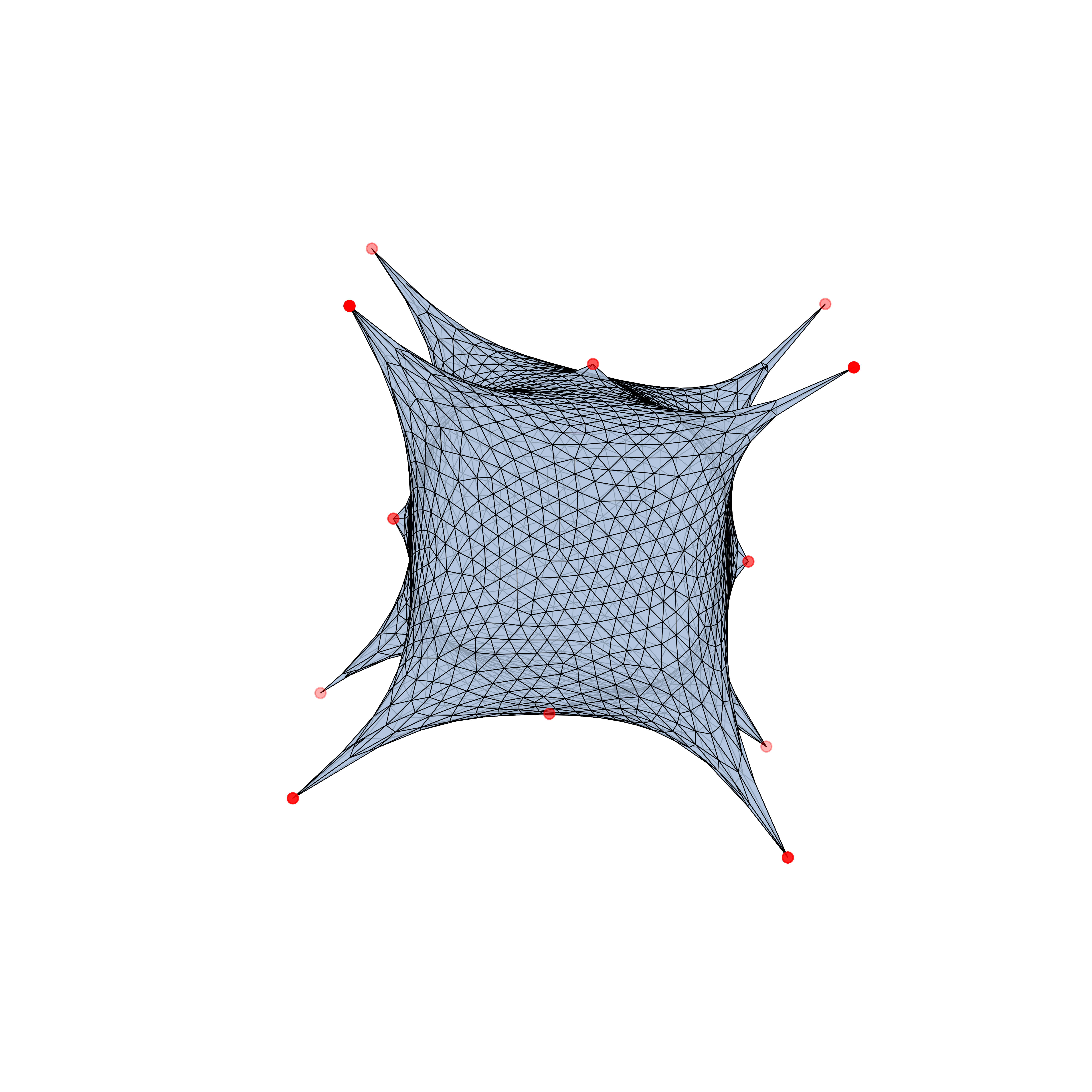}
        \caption{Mesh after 55 iterations.}
    \end{subfigure}
    \caption{Refinement and stabilization of the flexible foil mesh at iterations 40 and 55.}
    \label{fig:40_55_iterations}
\end{figure}

As the simulation progresses, the dynamic structure undergoes contraction and refinement due to 
the influence of elastic forces and adaptive constraints. Figures~\ref{fig:10_20_iterations} and 
\ref{fig:40_55_iterations} illustrate the iterative evolution of the mesh.

The evolution of the mesh follows the physical principles outlined in Section 2 (Mathematical Formulation). 
Initially, the mesh experiences rapid contraction forces due to pressure-driven deformation and the 
elastic response of the structure. This phase is dominated by high vertex displacement rates and uneven 
edge lengths, as the forces seek to minimize local distortions.

By iteration 20, the mesh has reached a more structured state, but irregularities remain in the edge 
length distribution. This phase correlates with the influence of adaptive snapping as described in 
Equation \eqref{eq:snapp}, 
which ensures that vertices gradually align with stable positions.

Between iterations 40 and 55, the mesh stabilizes into a near-equilibrium configuration, where elastic 
restoration forces and pressure forces reach a balance. The system behavior during this phase aligns with 
the convergence criteria outlined in Equation \eqref{eq:convergence}, where maximal displacement falls below the threshold \( \epsilon \).

\subsection{Numerical Stability and Damping Effects}

The numerical stability of the simulation is governed by two primary factors: damping control and 
CFL-based time step selection, as described in {Section 2.2.2 (Stability Considerations).

The damping coefficient \( c \) (introduced in Equation \eqref{eq:motion}) plays a crucial role in preventing excessive 
oscillations. An appropriate selection of \( c \) ensures:
\begin{itemize}
    \item Suppression of high-frequency oscillations that could introduce numerical artifacts.
    \item Faster convergence to equilibrium without underdamped oscillatory behavior.
    \item Preservation of structural consistency by limiting distortions.
\end{itemize}

Additionally, the time step \( \Delta t \) satisfies the CFL condition:

\begin{equation}
\Delta t < \frac{2}{\omega_{\text{max}}},
\end{equation}

where \( \omega_{\text{max}} \) is the highest eigenfrequency of the system, as detailed in Equation \eqref{eq:CFL}. 
Empirically, our simulation adheres to this constraint, preventing instability and divergence.

\subsection{Quantitative Convergence Analysis}

To further validate the stability of the simulation, we examine the maximum displacement and nearest-neighbor distances.

\begin{figure}[H]
    \centering
    \begin{subfigure}[t]{0.43\textwidth}
        \centering
        \includegraphics[width=\textwidth]{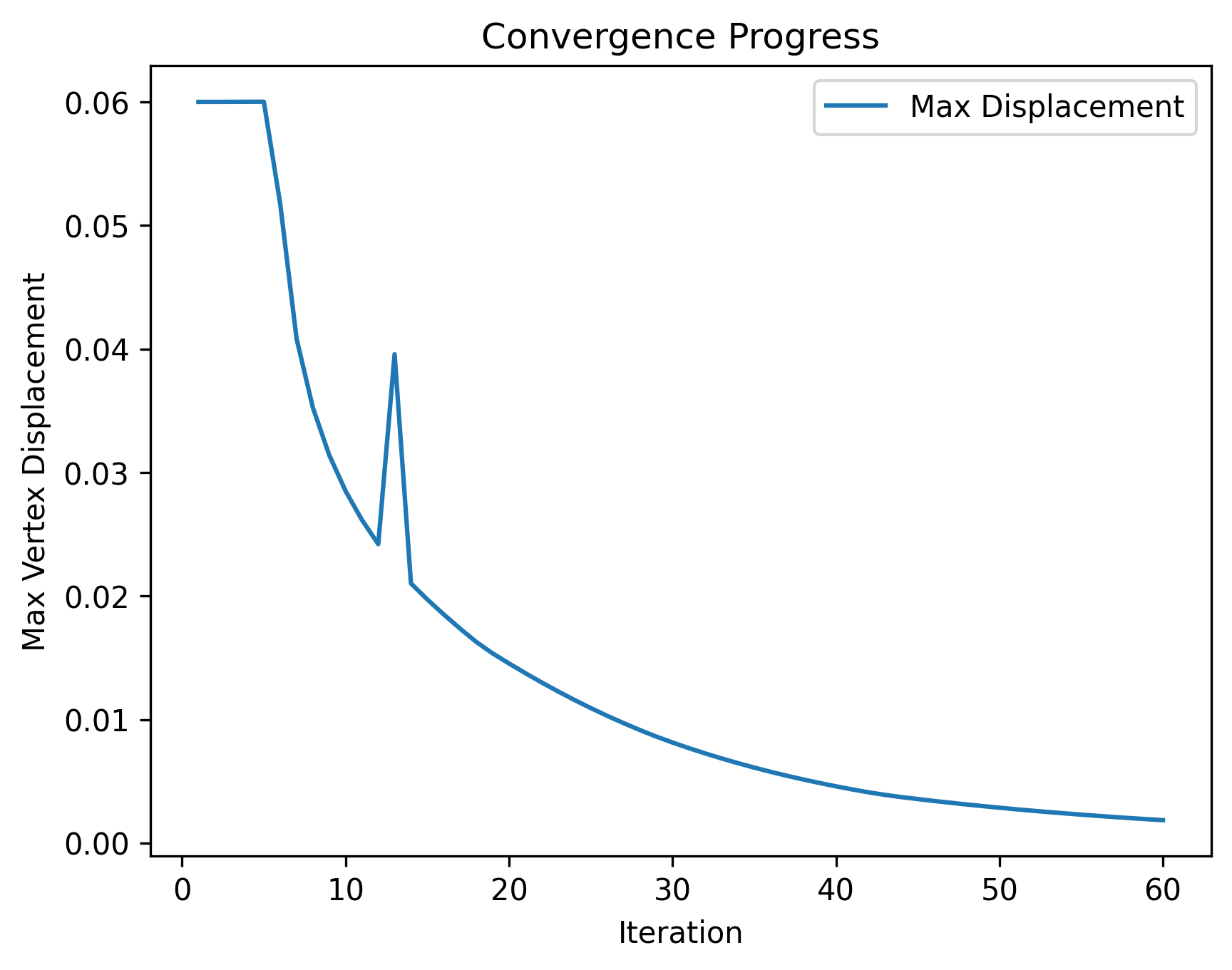}
        \caption{Maximal displacement during the iteration process.}
        \label{fig:CP}
    \end{subfigure}
    \hfill
    \begin{subfigure}[t]{0.48\textwidth}
        \centering
        \includegraphics[width=\textwidth]{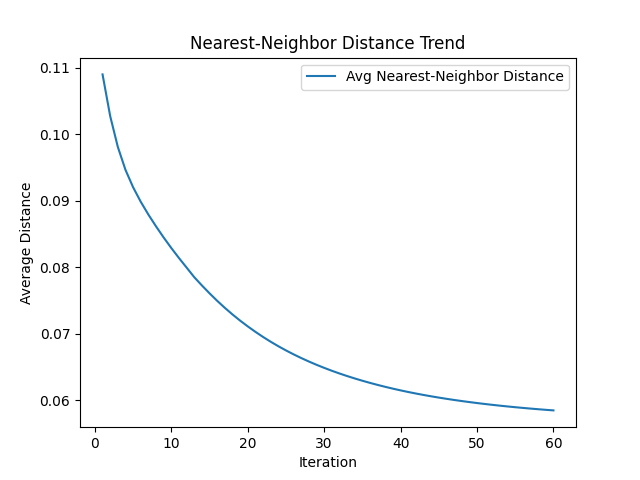}
        \caption{Average distance between mesh nodes.}
        \label{fig:NN}
    \end{subfigure}
    \caption{(Left) Maximal displacement of mesh points. (Right) Evolution of nearest-neighbor distances.}
\end{figure}

Figure~\ref{fig:CP} demonstrates that the maximum vertex displacement steadily decreases over iterations. 
The anomaly near iteration 13 aligns with the proximity effect of fixed vertices, where localized force 
imbalances temporarily introduce irregular displacement behavior. This is a direct consequence of the fixed 
vertex constraints 
described in Appendix A (Radius of Effectiveness).

Similarly, Figure~\ref{fig:NN} shows a progressive decrease in nearest-neighbor distances, indicating mesh 
densification as contraction progresses. This behavior validates the adaptive snapping mechanism and 
its effect on maintaining mesh uniformity.

\subsection{Summary of Findings}

The simulation results confirm the theoretical principles outlined in Section 2:
\begin{itemize}
    \item Mesh contraction is driven by pressure forces, with elastic forces ensuring structural integrity.
    \item Damping effects play a critical role in achieving numerical stability and smooth convergence.
    \item The CFL condition is maintained, preventing unstable behavior in explicit Euler integration.
    \item The snapping mechanism ensures that vertex distribution remains uniform over iterations.
\end{itemize}

These findings validate the flexible foil mesh approach as a stable and physically consistent method 
for constrained 3D mesh generation.

\section{Conclusion}

In this work, we investigated a flexible foil-based approach for mesh generation in confined 3D structures. 
The method integrates elasticity-driven deformation, pressure-induced contraction, and adaptive snapping 
constraints to iteratively refine an initial mesh while maintaining geometric consistency. The study explores 
the effects of these forces on mesh evolution and assesses numerical stability under explicit time integration.

\subsection{Summary of Findings}

The results demonstrate that the proposed approach can produce stable and well-structured meshes under 
spatial constraints. Several observations can be made based on the simulations:
\begin{itemize}
    \item The combination of elasticity and pressure forces enables a structured evolution of the mesh, 
    progressively reducing irregularities over iterations.
    \item The fixed vertex constraints play a critical role in guiding the deformation process, though their 
    placement influences local distortions in early iterations.
    \item The adaptive snapping mechanism improves mesh uniformity by dynamically adjusting vertex positions 
    based on local connectivity.
    \item The use of damping helps mitigate oscillations, leading to smoother convergence while ensuring compliance 
    with numerical stability conditions.
    \item The mesh evolves towards a near-equilibrium state, where force imbalances are minimized, 
    and vertex displacements stabilize.
\end{itemize}

These findings suggest that the approach is effective for generating well-structured meshes while preserving 
boundary constraints. However, as with any numerical method, certain limitations must be acknowledged.

\subsection{Limitations and Areas for Improvement}

While the proposed framework demonstrates promising behavior in mesh refinement, some aspects require further 
investigation:
\begin{itemize}
    \item The current implementation relies on explicit Euler integration, which imposes restrictions on 
    time step selection. Exploring alternative integration schemes may improve numerical stability and efficiency.
    \item The snapping mechanism is based on local connectivity but does not account for global shape optimization. 
    Further refinement strategies could improve overall uniformity.
    \item The method has been tested on relatively simple 3D structures. Extending the approach to more complex, 
    irregular geometries would provide additional insights into its robustness and applicability.
    \item The role of material properties and anisotropic effects has not been considered in this study. Incorporating 
    varying stiffness and heterogeneous material behavior could extend the method’s applicability.
\end{itemize}

\subsection{Final Remarks}

This study provides a preliminary exploration of flexible foil-based mesh generation under geometric constraints. 
The results indicate that elasticity, pressure, and adaptive snapping collectively contribute to a stable 
and structured mesh evolution. While the method shows promise, further refinement and validation are necessary 
to fully understand its applicability across different domains. Future work will focus on addressing the identified 
limitations and improving the framework to enhance its robustness and computational efficiency.

\newpage

\begin{appendices}
{\centering \LARGE \textbf{Supplementary Materials} \par}

\addcontentsline{toc}{section}{Supplementary Materials} 

\section{Radius of Effectiveness of Fixed Vertex Constraints}

In the simulation, the fixed vertices impose constraints on the neighboring mesh points. 
This influence diminishes as the distance from the fixed vertices increases. The radius of 
effectiveness quantifies how far this influence propagates before becoming negligible. 
Below, we describe the estimation process for this radius.

\subsection{Influence of Fixed Vertices}
The influence of fixed vertices is governed by two primary mechanisms:
\begin{itemize}
    \item \textbf{Scaling Function for Elastic Forces:} The elastic forces are scaled based on 
	the proximity of mesh points to fixed vertices. A decay function determines how stiffness 
	decreases with distance, controlled by the parameter $\mathit{distance\_factor\_strength}$.
    \item \textbf{Snapping Tolerance:} The snapping mechanism enforces alignment of mesh points 
	with fixed vertices if they are within a specified distance, defined by $\mathit{snapping\_tolerance}$.
\end{itemize}

\subsection{Radius of Effectiveness}
The radius of effectiveness, $R_e$, is the distance at which the influence of fixed vertices is considered negligible. It is estimated by considering the decay of stiffness and the snapping tolerance.

\paragraph{1. Stiffness Decay:}
The stiffness, $k_{\text{effective}}$, decays with distance $r$ from a fixed vertex as:
\[
k_{\text{effective}}(r) = k_{\text{base}} \cdot \left(1 + \mathit{distance\_factor\_strength} \cdot \frac{r}{d} \right)^{-1},
\]
where $d$ is the average nearest-neighbor distance. The radius of effectiveness is determined when $k_{\text{effective}}(r)$ falls to $1\%$ of its initial value:
\[
0.01 = \left(1 + \mathit{distance\_factor\_strength} \cdot \frac{R_e}{d} \right)^{-1}.
\]
Solving for $R_e$:
\[
R_e = \frac{d}{\mathit{distance\_factor\_strength}} \cdot (100 - 1).
\]
\[
R_e = \frac{99 \cdot d}{\mathit{distance\_factor\_strength}}.
\]

\paragraph{2. Snapping Tolerance:}
The snapping tolerance, $\mathit{snapping\_tolerance}$, provides an upper bound for the influence radius:
\[
R_e = \mathit{snapping\_tolerance}.
\]

\paragraph{3. Combined Influence:}
The effective radius of influence is the minimum of the decay-based radius and the snapping tolerance:
\[
R_e = \min\left(\frac{99 \cdot d}{\mathit{distance\_factor\_strength}}, \mathit{snapping\_tolerance}\right).
\]

\subsection{Number of Affected Neighbors}
The number of nearest neighbors affected, $N_e$, is estimated as:
\[
N_e \approx \frac{R_e}{d}.
\]
Substituting $R_e$:
\[
N_e \approx \min\left(\frac{99}{\mathit{distance\_factor\_strength}}, \frac{\mathit{snapping\_tolerance}}{d}\right).
\]

\end{appendices}

\newpage

\section*{Declarations}
All data-related information and coding scripts discussed in the results section are available from the 
corresponding author upon request.

\bibliographystyle{plain}

\begin{thebibliography}{99}

\bibitem{lee1980two}
Lee, D. T., and Schachter, B. J., "Two algorithms for constructing a Delaunay triangulation," \emph{International Journal of Computer and Information Sciences}, vol. 9, no. 3, pp. 219-242, 1980.

\bibitem{kazhdan2006poisson}
Kazhdan, M., Bolitho, M., and Hoppe, H., "Poisson surface reconstruction," in \emph{Proceedings of the Fourth Eurographics Symposium on Geometry Processing}, 2006, pp. 61-70.

\bibitem{preparata1985computational}
Preparata, F. P., and Shamos, M. I., \emph{Computational Geometry: An Introduction}, Springer, 1985.

\bibitem{Edelsbrunner1987}
H. Edelsbrunner, \textit{Algorithms in Combinatorial Geometry}, Springer, 1987.

\bibitem{MATLAB}
MathWorks, \textit{MATLAB Documentation: Boundary Function}, {https://www.mathworks.com/help/matlab/ref/boundary.html}.

\bibitem{botsch2010polygon}
Botsch, M., Kobbelt, L., Pauly, M., Alliez, P., and Lévy, B., \emph{Polygon Mesh Processing}, AK Peters/CRC Press, 2010.

\bibitem{provot1995deformation}
Provot, X., "Deformation constraints in a mass-spring model to describe rigid cloth behavior," in \emph{Graphics Interface}, 1995, pp. 147-154.

\bibitem{zienkiewicz1977finite}
Zienkiewicz, O. C., and Taylor, R. L., \emph{The Finite Element Method}, McGraw-Hill, 1977.

\bibitem{terzopoulos1987elastically}
Terzopoulos, D., Platt, J., Barr, A., and Fleischer, K., "Elastically deformable models," \emph{ACM SIGGRAPH Computer Graphics}, vol. 21, no. 4, pp. 205-214, 1987.

\bibitem{muller2007position}
M\"{u}ller, M., Heidelberger, B., Teschner, M., and Gross, M., "Position based dynamics," \emph{Journal of Visual Communication and Image Representation}, vol. 18, no. 2, pp. 109-118, 2007.

\bibitem{gingold1977smoothed}
R. A. Gingold and J. J. Monaghan, 
``Smoothed particle hydrodynamics: Theory and application to non-spherical stars,'' 
\textit{Monthly Notices of the Royal Astronomical Society}, vol. 181, no. 3, pp. 375--389, 1977.

\bibitem{brebbia1984boundary} 
Brebbia, C. A., Telles, J. C. F., and Wrobel, L. C. (1984). \textit{Boundary element techniques: theory and applications in engineering}. Springer-Verlag. 

\bibitem{cundall1979discrete} 
Cundall, P. A. and Strack, O. D. L. (1979). A discrete numerical model for granular assemblies. \textit{Géotechnique}, 29(1), 47-65.

\bibitem{sifakis2012fem}
Sifakis, E., and Barbic, J., "FEM simulation of 3D deformable solids: A practitioner's guide to theory, discretization, and model reduction," in \emph{ACM SIGGRAPH Courses}, 2012.

\bibitem{BaraffWitkin1998}
D. Baraff and A. Witkin.
\newblock Large steps in cloth simulation.
\newblock {\em Proceedings of SIGGRAPH}, pages 43--54, 1998.

\bibitem{taylor2008bio}
Taylor, Z. A., Cheng, M., and Ourselin, S., "High-speed nonlinear finite element analysis for surgical simulation using graphics processing units," \emph{IEEE Transactions on Medical Imaging}, vol. 27, no. 5, pp. 650-663, 2008.

\bibitem{TeranEtAl2005}
J. Teran, E. Sifakis, S. Blemker, V. Ng-Thow-Hing, C. Lau, and R. Fedkiw.
\newblock Creating and simulating skeletal muscle from the visible human data set.
\newblock {\em IEEE Transactions on Visualization and Computer Graphics}, 11(3):317--328, 2005.

\bibitem{bathe1996fluid}
Bathe, K. J., \emph{Fluid-Structure Interactions}, Springer, 1996.

\bibitem{debunne2001dynamic}
Debunne, G., Desbrun, M., Cani, M. P., and Barr, A. H., "Dynamic real-time deformations using space and time adaptive sampling," in \emph{Proceedings of SIGGRAPH}, 2001, pp. 31-36.

\bibitem{wang2015adaptive}
Wang, H., "Adaptive elasticity for real-time deformation," \emph{ACM Transactions on Graphics}, vol. 34, no. 4, pp. 1-12, 2015.

\bibitem{netzer2024form}
Moriya, N., "Form Convex Hull to Concavity: Surface Contraction Around a Point Set," \emph{arXiv preprint arXiv:2401.14189}, 2024.

\bibitem{Courant_1928} 
R. Courant, K. Friedrichs, and H. Lewy, \textit{Über die partiellen Differenzengleichungen der mathematischen Physik}, Mathematische Annalen, 1928.

\bibitem{Strikwerda_2004} 
J. C. Strikwerda, \textit{Finite Difference Schemes and Partial Differential Equations}, SIAM, 2004.

\bibitem{aistleitner2012}
C. Aistleitner, J. Brauchart, and J. Dick, 
\textit{Point sets on the sphere $\mathbb{S}^2$ with small spherical cap discrepancy}, 
\textit{Mathematics of Computation}, vol. 83, no. 290, pp. 2461--2480, 2014. 
Available at {https://arxiv.org/abs/1109.3265}.

\bibitem{extremelearning}
Extreme Learning, 
\textit{How to evenly distribute points on a sphere more effectively than the canonical Fibonacci lattice}, 2016. Available at: \\
{https://extremelearning.com.au/how-to-evenly-distribute-points-on-a-sphere-more-effectively-than-the-canonical-fibonacci-lattice/}

\bibitem{AmentaBern1999}
N. Amenta and M. Bern.
\newblock Surface reconstruction by Voronoi filtering.
\newblock {\em Discrete \& Computational Geometry}, 22(4):481--504, 1999.

\bibitem{bhatnagar2020machine}
Bhatnagar, B., Patel, P., and Teschner, M., "Machine learning-based approaches for deformable object simulation," \emph{Computer Graphics Forum}, vol. 39, no. 2, pp. 547-568, 2020.

\end{thebibliography}

\end{document}